\documentclass[twocolumn,preprintnumbers,superscriptaddress,amsmath,amssymb]{revtex4-2}
\usepackage{float}
\usepackage{graphicx}% Include figure files
\usepackage{dcolumn}% Align table columns on decimal point
\usepackage{bm}% bold math
\usepackage{epstopdf}
\usepackage{xcolor}
\usepackage{textcomp}
\usepackage{soul}% \st
\usepackage{csquotes}
\usepackage{amsbsy}
\usepackage{mathrsfs} % script-like, curvy letters.
\usepackage{amsmath}
\usepackage{bigints} % Big Integration
\usepackage{amsthm,amssymb} % Theorem Proof
\usepackage[utf8]{inputenc}
\usepackage[english]{babel}
\usepackage[shortlabels]{enumitem}
\usepackage{float}
\usepackage{mathrsfs,bigints,mathtools,dsfont}
\usepackage[colorlinks=true,linkcolor=blue,citecolor=blue]{hyperref}%
\usepackage[toc]{appendix}

\usepackage{textcmds}

\begin{document}
	
\title{{Self-organized bistability on globally coupled higher-order networks}}
%{\color{blue}Self-organized bistability in adaptive higher-order networks}}}
	
  \author{Md Sayeed Anwar}\affiliation{Physics and Applied Mathematics Unit, Indian Statistical Institute, 203 B. T. Road, Kolkata 700108, India}
   \author{Nikita Frolov} \affiliation{Laboratory of Dynamics in Biological Systems, Department of Cellular and Molecular Medicine, KU Leuven, Herestraat 49, 3000, Leuven, Belgium}
   \author{Alexander E. Hramov} \affiliation{Baltic Center for Neurotechnology and Artificial Intelligence, Immanuel Kant Baltic Federal University, 14, A. Nevskogo str., Kaliningrad 236016, Russia}
 \author{Dibakar Ghosh} %\email{dibakar@isical.ac.in}
 \affiliation{Physics and Applied Mathematics Unit, Indian Statistical Institute, 203 B. T. Road, Kolkata 700108, India}

 % REQUIRED

 \begin{abstract}
 Self-organized bistability (SOB) stands as a critical behavior for the systems delicately adjusting themselves to the brink of bistability, characterized by a first-order transition. Its essence lies in the inherent ability of the system to undergo enduring shifts between the coexisting states, achieved through the self-regulation of a controlling parameter. 
 %Current developments in the field of neuroscience underscore the significance of higher-order interactions among dynamic units and their pivotal role in molding collective behaviors.
 Recently, SOB has been established in a scale-free network as a recurrent transition to a short-living state of global synchronization. Here, we embark on a theoretical exploration that extends the boundaries of the SOB concept on a higher-order network (implicitly embedded microscopically within a simplicial complex) while considering the limitations imposed by coupling constraints. By applying Ott-Antonsen dimensionality reduction in the thermodynamic limit to the higher-order network, we derive SOB requirements under coupling limits that are in good agreement with numerical simulations on systems of finite size. We use continuous synchronization diagrams and statistical data from spontaneous synchronized events to demonstrate the crucial role SOB plays in initiating and terminating temporary synchronized events. We show that under weak coupling consumption, these spontaneous occurrences closely resemble the statistical traits of the epileptic brain functioning. 		
 \end{abstract}

 \maketitle	

 \section{Introduction} \label{intro}
 Multistability is a prevalent phenomenon observed in both man-made and real-world systems, characterized by the presence of various stable states that can be sustained under consistent conditions \cite{binder1987theory,binney1992theory}. This phenomenon plays a crucial role in regulating processes in living systems operating on different scales, from organ system interactions to neural synchronization \cite{angeli2004detection,pisarchik2014control,pigorini2015bistability}. In the context of consciousness, the switching ability between different local coherence patterns in the cortical region of the brain is essential \cite{koch2016neural}. Typically, in normal conditions, neural activity in the brain demonstrates distinct power-law (scale-free) distributed avalanches, which is indicative of underlying self-organized criticality (SOC) \cite{linkenkaer2001long,beggs2003neuronal,shew2013functional,bak1988self}. 
\par Nevertheless, in situations when epilepsy prevails \cite{lehnertz2009synchronization,luttjohann2015dynamics} or when inhibitory mechanisms are suppressed \cite{louzada2012suppress}, exceptionally large events occur, exceeding what would be anticipated under critical conditions. Furthermore, the size distributions of these events exhibit two distinct peaks (i.e., bimodal distribution), indicating the presence of some form of underlying bistable dynamics of neural ensembles. To explain this phenomenon, Di Santo et al. \cite{di2016self} introduced the concept of self-organized bistability (SOB) which is the counterpart of SOC, for systems adjusting themselves to the verge of bistability of a first-order phase transition. The switching between two stable states occurs through self-tuning of the control parameter influenced by driving force and dissipation. Their findings underscore the fundamental importance of the SOB theory in understanding complex neural dynamics, particularly in the context of epilepsy \cite{di2016self,buendia2020self}.
\par Researchers have started to take into account the networked organization of neural populations to get a deeper understanding of the genesis of bistable dynamics \cite{kalitzin2019epilepsy}. In this context, our group has recently reported a networked extension of SOB theory  \cite{frolov2021extreme,frolov2022self}, emphasizing the need to comprehend how structural characteristics drive SOB on complex networks, ultimately contributing to the occurrence of large avalanches. The proposed model is based on the assumption that the organization of the local neural populations is scale-free (SF) \cite{barabasi2003scale} and the connections between neurons are merely pairwise. The reason behind this specific choice was that the pairwise SF structure exhibits an explosive transition (bistability dynamics) under some circumstances such as degree-frequency correlations \cite{gomez2011explosive,coutinho2013kuramoto,boccaletti2016explosive}. 
\par However, the grouping of neural ensembles is not limited to only dyadic connections but also includes many-body (higher-order) interactions among the neurons (i.e., simultaneous interaction between more than two neurons) \cite{petri2014homological,giusti2016two,sizemore2018cliques,anwar2022stability,anwar2023neuronal}. Glial cells and astrocytes, in particular, are thought to be potential sources of higher-order interactions in the brain, as they are believed to modulate the synaptic interaction in groups of neurons \cite{cho2016optogenetic,tlaie2019high,fellin2004neuronal,allen2009glia}. Strikingly, the alteration in astrocyte activity is thought to have a role in the generation of epileptic seizures~\cite{seifert2010astrocyte, seifert2013neuron, diaz2019glia}. Recent theoretical efforts revealed an emergent synchronization in biological neurons interacting with glial cells supporting an established astrocyte-based view on the origin of epilepsy~\cite{makovkin2022controlling, makovkin2020astrocyte}. Following current advances in graph theory, one can encode such higher-order neuron-astrocyte interaction using a mathematical language of simplicial complexes~\cite{bianconi2021higher} and hypergraphs~\cite{berge1985graphs}. Such higher-order structures provide a more thorough understanding of complex systems governed by concurrent interactions of numerous agents~\cite{battiston2020networks,battiston2021physics,majhi2022dynamics,boccaletti2023structure}, of which epileptic brain is a perfect example.
\par Motivated by the above discussion and inspired by the principles of neuron-glial coupling, we introduce a conceptual model that replicates the activity of the epileptic brain and more broadly recapitulates the phenomenon of SOB in complex systems with higher-order interactions. This model inherits some structural properties of neuron-astrocyte interaction and can be considered as an extension of the networked model~\cite{frolov2022self} to the higher-ordered framework, where coupling constraint imposed on triads of network nodes is adapted and self-tuned similar to what is proposed in the SOB theory. An intriguing feature of the current model is its departure from the pairwise networked one in a distinct manner. Unlike the former, the present model does not presuppose that the arrangement of local neural populations follows an SF configuration. Instead, it embraces a notion of homogeneity in the organization of these neural populations and thus indicates that the emergence of SOB on networked systems might not always be determined by underlying specific structural properties. This is possible due to the added nonlinearities imposed by higher-order interactions to exhibit explosive synchronization transition (bistable collective behavior) without any additional constraints on the organization of underlying connection topology \cite{skardal2020higher,skardal2019abrupt,millan2020explosive,sun2023dynamic}. 
\par In alignment with the SOB theory and our previous result, this model has imitated the manifestation of epileptic seizure initiation and automatic cessation. This phenomenon arises from the intricate interplay of abrupt synchronization transition and coupling constraints within a supercritical synchronized state and is termed as \qq{exteme synchronization event} for its spontaneous emergence and fleeting duration. However, in contrast to the previous result, here, the distribution of return times (i.e., the time intervals between neighboring spontaneous and short-termed synchronous states) does not converge to a constant power law scaling exponent. Instead, this scaling exponent demonstrates a noteworthy trend of variation within the domain of self-organized behavior (SOB). Notably, this exponent becomes increasingly negative as the system undergoes evolution towards the forward transition into the coherent state. 
\par Here, we delve into the comprehensive theoretical analysis to deduce the conditions governing SOB within the proposed higher-order model while considering the limitations imposed by coupling constraints. As an underlying higher-order structure, we here consider a globally coupled simplicial complex with interactions up to three-body, which is indeed the simplest homogeneous higher-order structure to exhibit explosive transition without imposing any additional constraints \cite{skardal2020higher}. For this model, we determine the requirements of SOB under coupling restrictions by employing Ott-Antonsen dimensionality reduction \cite{ott2008low} in the thermodynamic $(N \to \infty)$ limit, which aligns well with numerical simulations conducted on finite-size systems. We illustrate the critical role played by SOB in producing and ending temporary synchronized events by looking at continuous synchronization diagrams and statistical characteristics of spontaneous synchronized events. We demonstrate that these events mimic the statistical characteristics seen in epileptic neural functionality under weak coupling consumption. 

 \section{ The model } \label{model}
	 To begin with, we consider an ensemble of $N$ globally coupled Kuramoto phase oscillators subject to two- and three-body interactions (i.e., a simplicial complex of dimension $2$), whose rotation is given by the following set of differential equations, 
	\begin{subequations} \label{model_eq}
        \begin{multline}\label{model_eq1}
            \dot{\theta}_{i} = \omega_{i} + \lambda_{i}
			\bigg[\frac{k_{1}}{N}\sum\limits_{j=1}^{N} \sin(\theta_{j}-\theta_{i}) +\\ +
			\frac{k_{2}}{N^2}\sum\limits_{j=1}^{N} \sum\limits_{k=1}^{N}
			\sin(2\theta_{j}-\theta_{k}-\theta_{i})\bigg],
        \end{multline}
		\begin{equation}\label{model_eq2}
			\dot{\lambda}_{i} = \alpha  (\lambda_{0}-\lambda_{i}) - \beta r.
		\end{equation}          		
	\end{subequations}
	In Eq.~\eqref{model_eq1}, $\theta_{i}$ and $\dot{\theta}_{i}$, $(i=1,2,\dots, N)$ are the instantaneous phase and velocity of each $i$th oscillator, $\omega_{i}$ is the natural (intrinsic) frequency assumed to be drawn from a distribution $g(\omega)$. $k_{1}$ and $k_{2}$ are the coupling strengths associated with two-body ($1$-simplex) interactions and three-body ($2$-simplex) interactions, respectively. Eq.~\eqref{model_eq2} accounts for the temporal behavior of each unit by their connection to resource bath $\lambda_{i}$ through a diffusive coupling. Eq.~\eqref{model_eq2} thus explains the self-tuning of individual coupling strength $\lambda_{i}$ in accordance with Di Santo et al.~\cite{di2016self}. This equation for resource in the form of Eq.~\eqref{model_eq2} was introduced in the previous work \cite{frolov2021extreme}. Here, the excitability consumption is considered as a function of order parameter $r=\dfrac{1}{N}|\sum\limits_{j=1}^{N}e^{i\theta_{j}}|$. The first term in the right-hand side of Eq. \eqref{model_eq2} describes the excitability recovery at a rate $\alpha $, while the second term attribute to resource constraint at a rate $\beta $. $\lambda_{0}$ defines the depth of individual resource bath that indicates the level of excitability in the absence of resource constraint. 

\subsection{Ott-Antonsen reduction}
\par To provide a deeper analytical insight into the underlying higher-order network dynamics given by Eqs. \eqref{model_eq1} and \eqref{model_eq2}, we employ the Ott-Antonsen dimensionality reduction formalism. We therefore, introduce the generalized complex order parameter $z^{(p)}=r^{(p)}e^{i\phi_{p}}=\frac{1}{N}\sum\limits_{j=1}^{N}e^{ip\theta_{j}}$, $p=1,2$, where $r^{(p)}$ and $\phi_{p}$ are the amplitude and argument of $p$-cluster order parameter. Here, $z^{(1)}$ is the conventional Kuramoto order parameter, while $z^{(2)}$ represents two-cluster order parameter~\cite{skardal2019abrupt}. Then, the evolution of Eq. \eqref{model_eq1} can be rewritten in terms of the complex order parameters $z^{(1)}$ and $z^{(2)}$ as,   
	\begin{equation}\label{complex_model}
		\begin{array}{l}
			\dot{\theta}_{i}=\omega_{i}+ \frac{\lambda_{i}}{2\mathrm{i}}\big[He^{-\mathrm{i}\theta_{i}}-H_{c}e^{\mathrm{i}\theta_{i}}\big],\\
			 H=k_{1}z^{(1)}+k_{2}z^{(2)}z^{(1)}_{c}, 
		\end{array}
	\end{equation}
    where $z^{(1)}_{c}$ indicates the complex conjugate of $z^{(1)}$. To move forward, in Eq.\eqref{complex_model}, we propound $\lambda_{i}=\lambda$. We then consider the continuum limit $N\to \infty$ where the state of the system can be represented by a continuous density function $f(\theta,\omega,t)$ such that at time $t$, $f(\theta,\omega,t)\delta\theta\delta\omega$ describes the density of oscillators with the intrinsic frequency between $\omega$ and $\omega+\delta \omega$, and phases lying in the interval $[\theta, \theta+\delta\theta]$. The density function $f(\theta,\omega,t)$ satisfies the normalization condition $\int_{0}^{2\pi} f(\theta,\omega,t)d\theta=1$ and moreover because the number of oscillators in the system remains reserved, the density function must satisfy the continuity equation,
    \begin{equation} \label{continuity}
    	\begin{array}{l}
    		\dfrac{\partial f(\theta,\omega,t)}{\partial t}+\dfrac{\partial}{\partial \theta} [f v(\theta,\omega,t)] = 0.
    	\end{array}
    \end{equation}
    Besides the order parameters can be expressed in terms of density function as $z^{(p)}=\int\int e^{ip\theta}f(\theta,\omega,t)d\theta d\omega$. Now, since each oscillator's intrinsic frequency is fixed (drawn from a distribution $g(\omega)$) and $f(\theta,\omega,t)$ is $2\pi$-periodic with respect to $\theta$, the density function can be expanded into Fourier series of the form   
	\begin{equation} \label{fourier}
		\begin{array}{l}
			f(\theta,\omega,t)=\frac{1}{2\pi}\bigg(1+\sum\limits_{n=1}^{\infty}[a_{n}(\omega,t)e^{in\theta}+c.c]\bigg),
		\end{array}
	\end{equation} 
    where $a_{n}(\omega,t)$ is the $n$th Fourier coefficient, and c.c accounts for the complex conjugate of the previous sum. We then follow the Ott-Antonsen hypothesis that the Fourier coefficients decay geometrically and the sum converges, i.e., $a_{n}(\omega,t)=[a(\omega,t)]^{n}$, where $a(\omega,t)$ is analytic in complex $\omega$ plane and $|a(\omega,t)| \ll 1$ which is necessary for the convergence of the series. Thereafter, inserting the ansatz into the expression of $f$ and $f$ into the continuity equation \eqref{continuity} along with the expression for $v=\dot{\theta}$ [Eq.\eqref{complex_model}], all Fourier modes fall onto the   
	same constraint for $a(\omega,t)$, satisfying a single differential equation
	\begin{equation}\label{reduced_eq}
		\begin{array}{l}
			\dfrac{\partial a}{\partial t} = -ia\omega + \dfrac{\lambda}{2} \{H_{c}-
			Ha^{2}\}.
		\end{array}
	\end{equation}
  Moreover, the order parameter in the thermodynamic limit becomes
    \begin{equation}
	   \begin{split}\label{thermodynamic_op}
		z^{(m)} &= \int\int e^{im\theta}f(\theta,\omega,t)d\theta d\omega \\ &= \int_{-\infty}^{\infty} d\omega g(\omega) a_{c}^{(m)} (\omega,t).
	   \end{split} 
    \end{equation}
Now in order to obtain the critical synchronization transition, we analyze the stability of incoherent state $f(\theta,\omega,t)=\frac{1}{2\pi}$. In this regard, $a(\omega,t)=0$ is always a trivial solution of Eq. \eqref{reduced_eq} which corresponds to the incoherent state $f(\theta,\omega,t)=\frac{1}{2\pi}$ in Eq. \eqref{fourier}. Linearizing Eq. \eqref{reduced_eq} about the solution $a(\omega,t)=0$, one can obtain the following linear equation in terms of the perturbed density $\zeta(\omega,t)$,	
	\begin{equation}\label{perturbed_eq}
		\begin{array}{l}
			\dfrac{\partial \zeta}{\partial t} + i \zeta \omega = \dfrac{\lambda
				k_{1}}{2} \bigintss_{-\infty}^{\infty} d\omega g(\omega) \zeta(\omega,t).
		\end{array}
	\end{equation}
Now, we seek a solution of the form, $\zeta(\omega,t)=\zeta_{0}e^{\mu t}$, where $\mu$ is the eigenvalue of	Eq. \eqref{perturbed_eq} and is independent of $\omega$. Inserting the above expression for $\zeta(\omega,t)$ in equation \eqref{perturbed_eq}, we obtain the reduced equation as follows,
	\begin{equation}\label{perturbed_eq2}
		\begin{array}{l}
			\zeta_{0}(\omega) \mu + i \zeta_{0} (\omega) \omega = \dfrac{\lambda
				k_{1}}{2} \bigintss_{-\infty}^{\infty} d\omega g(\omega) \zeta_{0} (\omega).
		\end{array}
	\end{equation}
To solve Eq. \eqref{perturbed_eq2}, we first start by denoting $B = \bigintss_{-\infty}^{\infty} d\omega g(\omega) \zeta_{0} (\omega)$. Then 
from Eq. \eqref{perturbed_eq2}, $\zeta_{0}(\omega)$ may be solved as:	
	\begin{equation}
		\begin{array}{l}
			\eta_{0}(\omega)  = \dfrac{\lambda k_{1} }{2} \dfrac{B}{\mu+i \omega}.
		\end{array}
	\end{equation}
By substituting this back into the expression for $B$, one eventually obtains,
	\begin{equation} \label{complex_critical_eq}
		\begin{array}{l}
			\dfrac{2}{\lambda k_{1}} = \bigintss_{-\infty}^{\infty} d\omega \dfrac{g(\omega)}{\mu + i \omega}.
		\end{array}
	\end{equation}
Note that Eq. \eqref{complex_critical_eq} explicitly accounts for the relation between eigenvalue $\mu$ and the coupling strengths $\lambda,k_{1}$. Now if $g(\omega)$ is an even function (e.g., Lorentzian, Gaussian frequency distributions), then Eq. \eqref{complex_critical_eq} transforms into the following form: 
   \begin{equation} \label{real_critical_eq}
   	\begin{array}{l}
   		\dfrac{2}{\lambda k_{1}} = \bigintss_{-\infty}^{\infty} d\omega \dfrac{\mu g(\omega)}{\mu^2 +  \omega^2}.
   	\end{array}
   \end{equation}
Now to obtain the critical coupling $\lambda_{*}$ which represents the transition from incoherent to coherent state, we consider the limit $\mu \to 0^{+}$ in Eq. \eqref{real_critical_eq}. Then $\dfrac{\mu}{\mu^2 + \omega^2}$ becomes more and more sharply peaked about $\omega=0$, but still the integral over $-\infty<\omega<\infty$ remains equal to $\pi$ \cite{strogatz2000kuramoto}. So, in the $\mu \to 0^{+}$ limit, Eq. \eqref{real_critical_eq} tends to   
	\begin{equation} \label{final_critical_eq}
		\begin{array}{l}
			\dfrac{2}{\lambda_{*} k_{1}} = \pi g(0).
		\end{array}
	\end{equation}
Therefore, for standard Gaussian intrinsic distribution $g(\omega)=\dfrac{1}{\sqrt{2\pi}}e^{-\frac{\omega^2}{2}}$, the critical coupling for the transition to synchrony is $\lambda_{*}=\frac{2}{k_{1}}\sqrt{\frac{2}{\pi}}$. Similarly for Lorentzian frequency distribution $g(\omega)=\dfrac{\Delta}{\pi(\omega^2+\Delta^2)}$ (with zero mean and half width $\Delta$), the required critical coupling for synchronization transition is $\lambda_{*}=\frac{2\Delta}{k_{1}}$.
\par Now, in order to evaluate the critical synchronization transition in the proposed system \eqref{model_eq1} and \eqref{model_eq2}, we must consider the influence of coupling constraint in Eq. \eqref{reduced_eq} and consequently to the latter equations. In the steady state, we have, 
\begin{equation}\label{steady_lambda}
		\begin{array}{l}
			\lambda= \lambda_{0}- \dfrac{\beta}{\alpha} r^{(1)},
		\end{array}
\end{equation}
since both $r$ and $r^{(1)}$ quantify the conventional Kuramoto order parameter. Hence by taking the coupling constraint into consideration and using Eqs. \eqref{final_critical_eq} and \eqref{steady_lambda}, we can define the critical transition point for our proposed model \eqref{model_eq1} and \eqref{model_eq2} as,   	 
	\begin{equation} \label{critical_point}
		\begin{array}{l}
			\lambda^{*}_{0}= \dfrac{2}{\pi k_{1}}g(0) + \dfrac{\beta}{\alpha}r^{(1)}_{0}, 
		\end{array}
	\end{equation}
	where $r^{(1)}_{0}= \dfrac{\eta}{\sqrt{N}}$, for some real $\eta$ is a finite estimation of the order parameter in the incoherent state~\cite{strogatz2000kuramoto,strogatz2004syncbook}.
	
\subsection{Solution of coherence}
We can further evaluate the advancement of order parameters for the choice of Lorentzian frequency distribution $g(\omega)=\dfrac{\Delta}{\pi(\omega^2+\Delta^2)}$. In this regard, the order parameter in Eq.~\eqref{thermodynamic_op} can be obtained using Cauchy's residue theorem by closing the contour to an infinite-radius semicircle in the negative-half complex plane, which results in  $z^{(1)}= a_{c}(-i\Delta,t)$ and $z^{(2)}=a_{c}^{2}(-i\Delta,t)=(z^{(1)})^{2}$. Thereafter, estimating Eq. \eqref{reduced_eq} at $\omega=-i\Delta$ and taking the complex conjugate, we eventually obtain
    \begin{multline}\label{coherent_eq1}
        2\dot{z}^{(1)}= 2i\omega z^{(1)} -2\Delta z^{(1)} +\lambda [ k_{1} z^{(1)}
        + k_{2} (z^{(1)})^{2} z^{(1)}_{c} \\ 
        - \{k_{1}z^{(1)}_{c} + k_{2} (z^{(1)}_{c})^{2}z^{(1)}\} (z^{(1)})^{2} ].
    \end{multline}
	Now, inserting $z^{(1)}=r^{(1)}e^{i\phi_{1}}$ and equating the real and imaginary parts of both sides of the equation yields
	\begin{subequations}
		\begin{equation} \label{coherent_eq2}
			2\dot{r}^{(1)} + 2\Delta r^{(1)} = \lambda r^{(1)} [1-(r^{(1)})^{2}] [k_{1}+k_{2}(r^{(1)})^{2}],
		\end{equation}
	 \begin{equation}
	 	 \dot{\phi_{1}}=0.
	 \end{equation}
	\end{subequations}
$r^{(1)}=0$ is always a solution of Eq. \eqref{coherent_eq2} whose stability is not affected by the presence of higher-order interaction. Nevertheless, the nonlinear terms which arise from the higher-order interactions arbitrate the likelihood of synchronized states. Particularly, one or two synchronous states exist, given by 	
	\begin{equation} \label{stable_unstable_op}
		\begin{array}{l}
			r^{(1)}_{\pm}= \sqrt{\dfrac{(k_{2}-k_{1})\pm \sqrt{(k_{1}+k_{2})^{2}-8k_{2}\dfrac{\Delta}{\lambda}}}{2k_{2}}},
		\end{array}
	\end{equation}
where $r^{(1)}_{+}$ ($r^{(1)}_{-}$) accounts for a stable (unstable) synchronous solution, subject to their existence. Note that Eq. \eqref{stable_unstable_op} is not enough to provide the solution of coherence of our considered model \eqref{model_eq} in the presence of coupling constraints. Therefore, in order to incorporate the influence of coupling constraint on the solution of coherence, we need to substitute the steady state relation \eqref{steady_lambda} in Eq. \eqref{stable_unstable_op}. Consequently, solving the Eqs. \eqref{stable_unstable_op} and \eqref{steady_lambda} together for the order parameter $r^{(1)}$ provides the solution of coherence for our considered model given by Eq. \eqref{model_eq}.

\par Now, since the essence of SOB lies in the fact that the underlying system must exhibit a first-order transition (bistable behavior) in the absence of coupling constraint \cite{di2016self}, we investigate the coupling condition under which the system exhibits bistable dynamics without coupling consumption. Thus, when $\beta=0$, one can simply substitute $\lambda_{0}$ in place of $\lambda$ in the expression of synchronous solution $r^{(1)}$ given by Eq. \eqref{stable_unstable_op} and eventually obtain
 \begin{equation} \label{stable_unstable_op2}
		\begin{array}{l}
			r^{(1)}_{\pm} \big \lvert_{\beta=0}= \sqrt{\dfrac{(k_{2}-k_{1})\pm \sqrt{(k_{1}+k_{2})^{2}-8k_{2}\dfrac{\Delta}{\lambda_{0}}}}{2k_{2}}}.
		\end{array}
	\end{equation}
Clearly, from the above expression we can obtain that the order parameter $r^{(1)}$ bifurcates from zero, i.e., the forward transition occurs at $\lambda^{f}_{0}=\dfrac{2\Delta}{k_{1}}$. In the case of backward transition, the stable and unstable solutions coexist at the interval of the bistable domain and collide with one another at the critical point of backward transition. So, using the condition $r^{(1)}_{+}=r^{(1)}_{-}$ in Eq. \eqref{stable_unstable_op2} we obtain the critical coupling for the backward transition as $\lambda^{b}_{0}=\dfrac{8k_{2}\Delta}{(k_{1}+k_{2})^{2}}$, which eventually gives us the condition $k_{2}>k_{1}$ for the existence of both the stable and unstable solution in the underlying system without any coupling consumption. Thus, in the absence of higher-order interactions $(k_{2}=0)$, or for $k_{2}\leq k_{1}$, the system will not exhibit a first-order transition. Throughout the study, we will, therefore use the coupling condition $k_{2}>k_{1}$ for pairwise and higher-order interactions to investigate the SOB behavior.

\begin{figure}[ht] 
  	\centerline{
  		\includegraphics[scale=0.26]{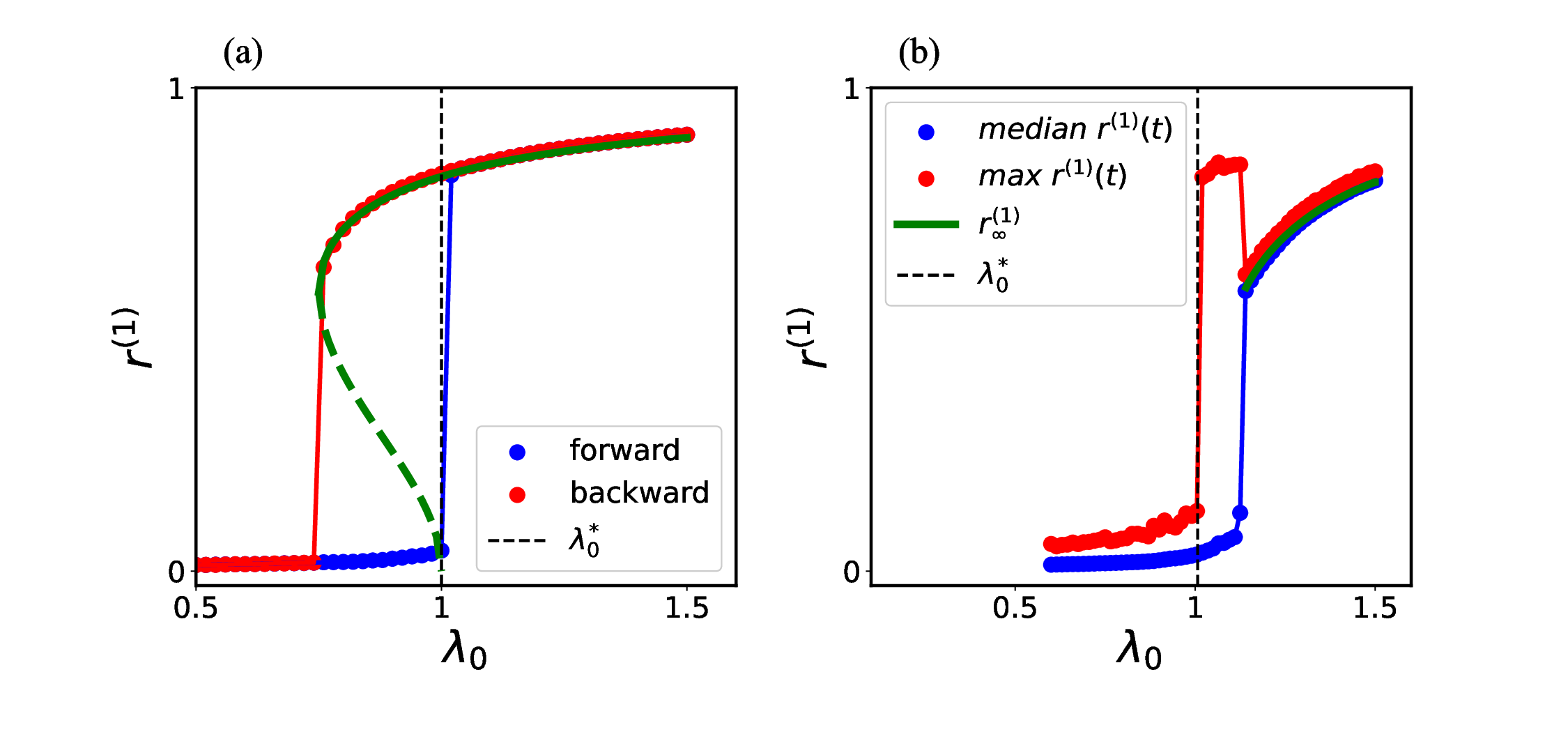}}
  	\caption{{\bf Lorentzian intrinsic frequency}. (a) Forward (blue circle) and backward (red circle) synchronization transition in the absence of excitability constraint $\beta $ (i.e., $\beta =0$), (b) Forward synchronization transition in the presence of excitability constraint $\beta =0.002$. The excitability recovery rate for both panels is fixed at $\alpha =0.003$. In panel (a), the forward and backward synchronization diagram and in panel (b), the maximum (red circle) and median value (blue circle) of order parameter $r^{(1)}$ are evaluated over a long time interval of $3\times 10^{4}$ time units after an initial transient period of $10^{3}$ time units. In (a), the vertical dashed black line corresponds to the critical synchronization transition $\lambda^{*}_{0}$, given by Eq. \eqref{critical_point} and the solid green curves represent analytical curves of the order parameter in the coherent state, while the dashed green line corresponds to the unstable solution in the absence of excitability constraint, obtained by solving the Eq. \eqref{stable_unstable_op2} and for (b) the solid green line delineates the coherent solution obtained using both Eqs. \eqref{stable_unstable_op} and \eqref{steady_lambda}, respectively. Here in the figure legend $r^{(1)}_{\infty}$ has been used to indicate the solution of coherence at $N\to \infty$ limit. }
  	\label{lor_beta0_beta_002}
  \end{figure}

\section{ Results} \label{numerical}
To illustrate our findings, we start with integrating Eqs. \eqref{model_eq1} and \eqref{model_eq2} for $N=10^{4}$ units using fourth-order Runge-Kutta scheme subject to adaptive time-stepping. Throughout the main text, we will restrict ourselves to the results associated with Lorentzian intrinsic frequency distribution. The investigation with Gaussian frequency distribution is illustrated in Appendix~\ref{app:gaussian}, which reflects qualitatively similar results.    

%\subsection{Lorentzian intrinsic frequency}
 \par  Therefore, we consider the scenario where the intrinsic frequencies are drawn from the standard Lorentzian distribution with zero mean and half-width $\Delta=0.5$. Considering the pairwise and three-body coupling strength to be fixed at $k_{1}=1$ and $k_{2}=3$ (as it satisfies our coupling condition), we plot the Kuramoto order parameter $r^{(1)}$ as a function of resource depth $\lambda_{0}$, as $\lambda_{0}$ is first increased from $\lambda_{0}=0$ adiabatically to an adequately large value and then decreased back. Figure~\ref{lor_beta0_beta_002} demonstrates the results for two different values of excitability constraint $\beta=0$ (panel (a)) and $\beta=0.002$ (panel (b)) at fixed excitability recovery rate $\alpha =0.003$. In the absence of excitability constraint $\beta $, i.e., $\beta =0$, our system mimics the Kuramoto model with higher-order interactions~\cite{skardal2020higher} and shows an abrupt transition to synchronization with associated hysteresis loop, also predicted from the expression of the synchronous solution~\eqref{stable_unstable_op2} [see Fig.~\ref{lor_beta0_beta_002}(a)].
 
\par In the presence of excitability constraint $\beta =0.002$ (i.e., with the coupling consumption), it can be observed that the forward transition to synchronization is delayed with respect to the critical transition point $\lambda_{0}^{*}$. Thus, Fig.~\ref{lor_beta0_beta_002}(b) reports that both the synchronized and desynchronized solutions are conceivable in the interval between $\lambda_{0}$ and the commencement of forward transition. Now according to Eq.~\eqref{model_eq2}, a stable synchronized state can only exist if the coupling dissipation of each individual is offset by their recovery, i.e., $\alpha (\lambda_{0}-\lambda_{i})\ge \beta r$. Therefore, in the region between $\lambda_{0}^{*}$ and the onset of the synchronization transition, the incoherent state $r^{(1)} \approx 0$ is most desirable by the system because the latter requirement is not met here. However, sudden shifts from the incoherent state to the coherent state occur within this domain, as can be observed from the maximum value of the order parameter $r^{(1)}(t)$.

\par Following the results of Fig.~\ref{lor_beta0_beta_002}(b), three distinctive regions can be identified under the coupling consumption: (i) $\lambda_{0}<\lambda_{0}^{*}$, where the incoherent state (i.e., $r^{(1)} \approx 0$) is the only possible dynamics. We call this domain the region of subcritical dynamics; (ii) The region between the critical synchronization transition point $\lambda_{0}^{*}$ and the onset of forward transition, called the region of critical bistable dynamics. Here we can find both coherent and incoherent dynamics, but the incoherent dynamics is most desirable; (iii) the region of supercritical dynamics corresponds to a coherent dynamics, obtained for the values of $\lambda_{0}$ larger than the value associated with the forward transition. 
  \begin{figure}[ht] 
  	\centerline{
  		\includegraphics[scale=0.39]{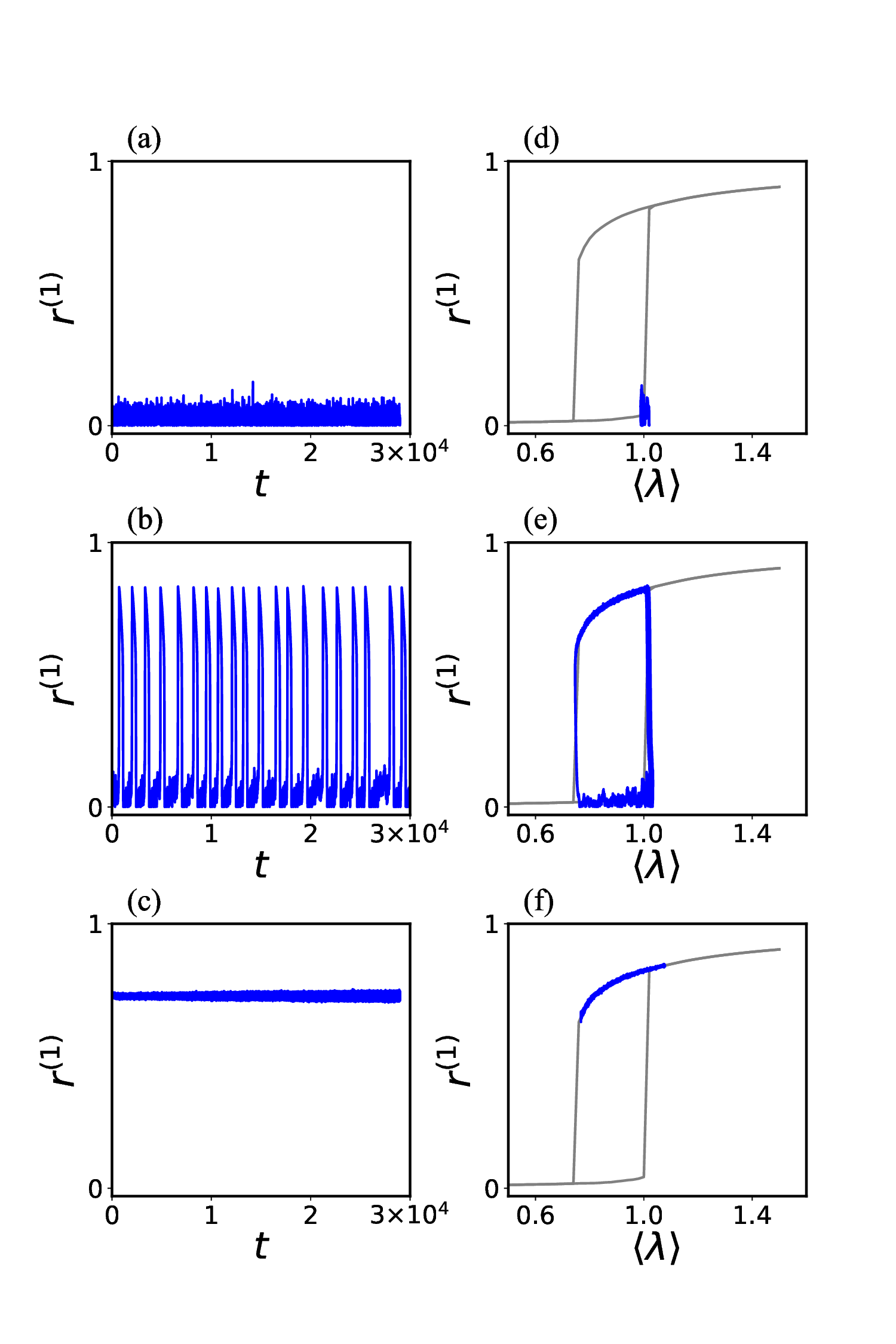}}
  	\caption{{\bf Lorentzian intrinsic frequency}. Macroscopic dynamics of higher-order globally coupled network. The left and right columns represent the long-term time dependency of global order parameter $r^{(1)}$ and corresponding phase portrait on the $(r^{(1)},\langle \lambda \rangle)$ plane respectively, for fixed values of $\alpha =0.003$, $\beta =0.002$ and different excitability bath depth $\lambda_{0}$. In the top, middle, and bottom rows, the value of excitability bath depth is respectively, $\lambda_{0}=0.9$, $1.08$, $1.3$. Dashed gray lines in (d), (e), and (f) depict the forward and backward synchronization diagrams $r^{(1)}(\lambda)$ in the absence of excitability constraint.}
  	\label{lor_ts}
  \end{figure}
  
\par To illustrate these dynamics in terms of the macroscopic parameters under the variation of $\lambda_{0}$, we plot in Fig.~\ref{lor_ts} the time series of the order parameter $r^{(1)}(t)$ and the accompanying phase portrait on the $(r^{(1)},\langle \lambda \rangle)$ plane for typical values of $\lambda_{0}$ with $\alpha =0.003$ and $\beta =0.002$, where $\langle \lambda \rangle=\frac{1}{N}\sum\limits_{j=1}^{N} \lambda_{j}$ indicates the ensemble average of the coupling ability. As anticipated, for $\lambda_{0}=0.9$ that belongs to the region of subcritical dynamics, the system converges at the lower branch of the hysteresis loop, corresponds to the incoherent solution $r^{(1)} \approx 0$  [Figs.~\ref{lor_ts}(a) and \ref{lor_ts}(d)]. In the domain of critical bistability, the system acquires stable incoherent and unstable coherent dynamics, and therefore sudden shifts from a desynchronized state to a synchronized state can be observed [Fig.~\ref{lor_ts}(b)]. Due to the presence of sufficient excitability resources, the system abruptly departs from the stable state and moves toward the unstable state. However, the resource is inadequate to maintain a position close to the unstable state and returns back to the stable one. This interesting dynamics is also clear from the phase trajectories in Fig.~\ref{lor_ts}(e). One can observe that the system mostly stays near the stable incoherent state on the lower branch of the hysteresis loop near the critical point $\lambda_{0}^{*}$. However, the unstable drifting occasionally causes the trajectory to move toward the upper branch of the hysteresis loop. As soon as the trajectory reaches the top branch, resource consumption forces a backward transition. The system goes through the backward transition as the coupling resource runs out. The trajectory eventually returns to its starting location close to the forward transition point on the bottom branch of the hysteresis loop, thanks to the diffusive process that enables coupling recovery and this process repeats over time. Lastly, for a large value of $\lambda_{0}$ (i.e., in the supercritical domain), due to a large amount of resource consumption, the system converges to the state of synchronized dynamics and stays on the upper branch of the hysteresis loop [Figs.~\ref{lor_ts}(c) and \ref{lor_ts}(f)]. 
\begin{figure}[t] 
  	\centerline{
  		\includegraphics[scale=0.26]{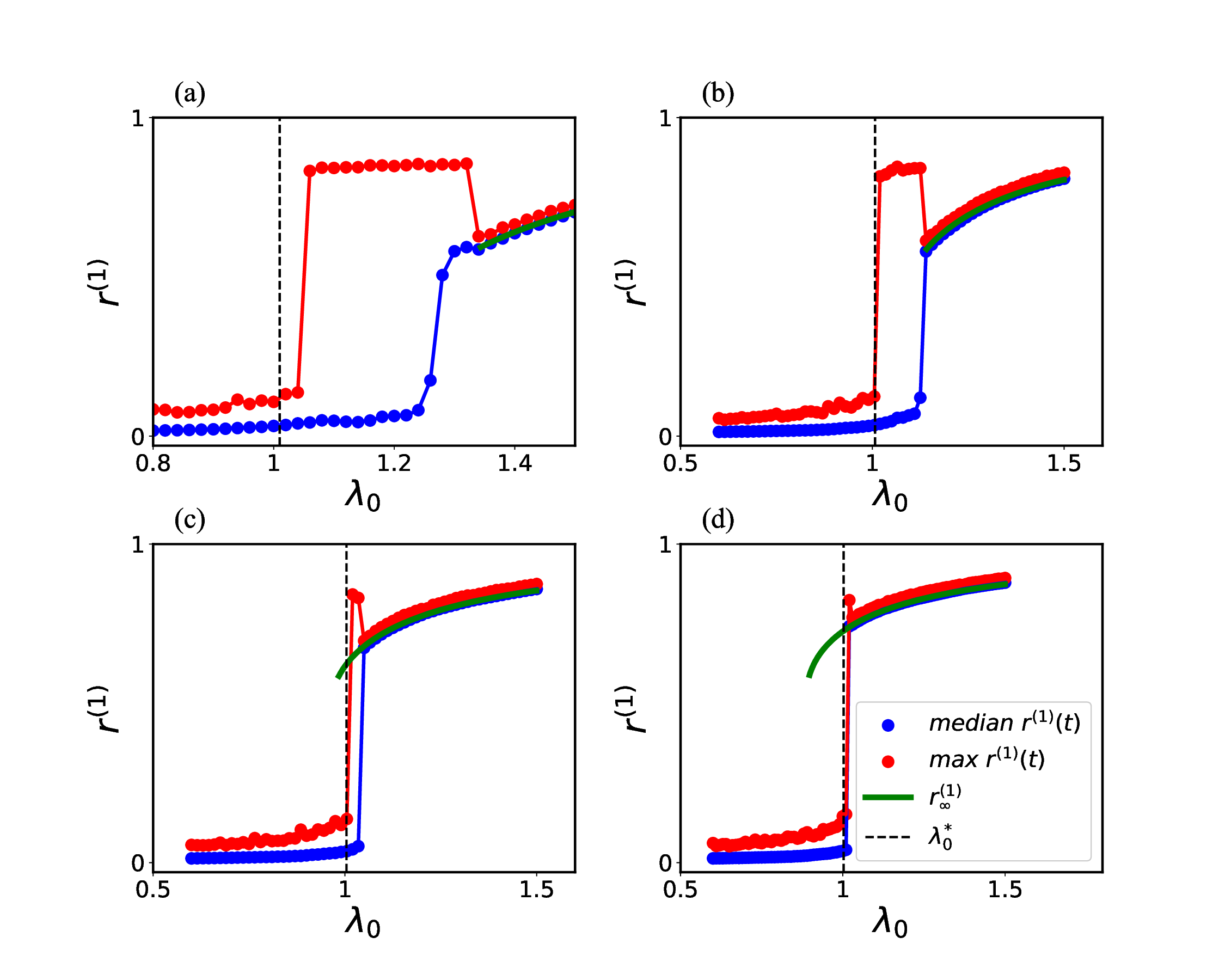}}
  	\caption{{\bf Lorentzian intrinsic frequency}. Forward synchronization transition in terms of $r^{(1)}(\lambda_{0})$ for fixed value of excitability constraint $\beta =0.002$ and different values of excitability recovery rate $\alpha $: (a) $\alpha =0.002$, (b) $\alpha =0.003$, (c) $\alpha =0.005$, and (d) $\alpha =0.008$. The solid green curves in each panel display the analytical curves of the order parameter in the coherent state obtained from Eqs. \eqref{steady_lambda} and \eqref{stable_unstable_op}, respectively. The vertical dashed black lines correspond to the critical synchronization transition $\lambda_{0}^{*}$, given by Eq. \eqref{critical_point}.}
  	\label{lor_different_alpha}
  \end{figure} 
\par Now, following the Eqs.~\eqref{steady_lambda} and \eqref{stable_unstable_op}, we can obtain that the dynamics of the system is less affected by the impact of coupling constraints when the value of consumption rate $\beta $ is being decreased, or equivalently the recovery rate $\alpha $ is increased. Therefore, to scrutinize how the relationship between $\beta $ and $\alpha $ impacts the crucial dynamics of the considered model in Fig.~\ref{lor_different_alpha} we plot the variation of order parameter $r^{(1)}$ as a function of $\lambda_{0}$ for different values of recovery rate $\alpha $ and fixed coupling consumption rate $\beta =0.002$. As expected, it can be observed that as the recovery rate $\alpha $ increases, the domain exhibiting the critical bistability behavior, which is bounded by the beginning of the forward transition on the right and critical coupling $\lambda_{0}^{*}$ on the left, gradually decreases in size. Eventually, the transition to coherence becomes abrupt. Further solving the Eqs.~\eqref{steady_lambda} and \eqref{stable_unstable_op}, we plot the curves of the stable synchronized state (solid green curves), which shows that with increasing $\alpha $ the curve of the stable coherent state extends to the left and crosses the critical transition point for higher values $\alpha$. This characterizes that the impact of coupling constraint decreases with increasing $\alpha $, and as a result for larger values of $\alpha $ an abrupt transition to the coherent state emerges with an associated hysteresis loop. Therefore, our findings for a finite-size system are in excellent agreement with the results obtained in the thermodynamic limit $(N \to \infty)$.
  \begin{figure*}[hpt] 
  	\centerline{
  		\includegraphics[scale=0.85]{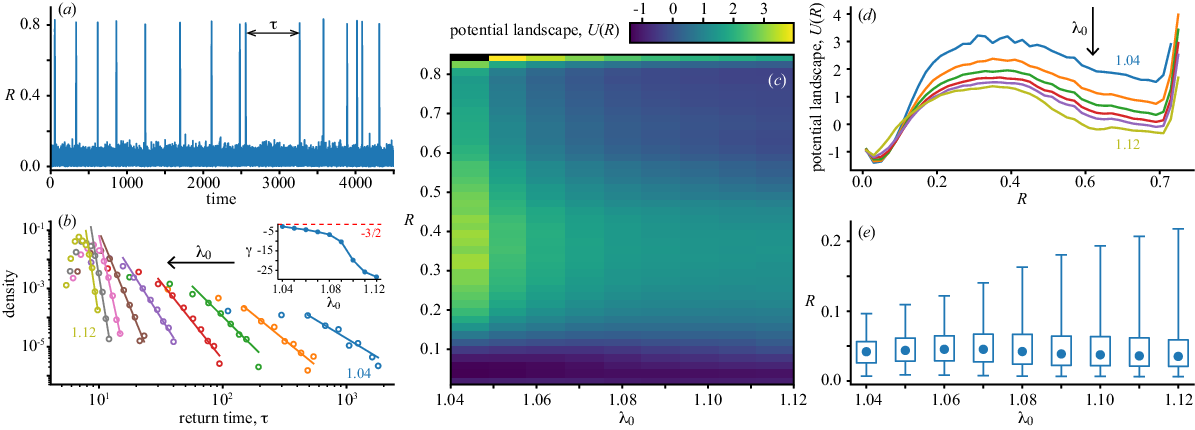}}
  	\caption{{\bf Analysis of the return times for $\alpha =0.003$ and $\beta =0.002$}. (a) Typical time-series of the order parameter $R(t)$ for $\lambda_0=1.04$. Here, return time $\tau$ is indicated with a double-headed arrow. (b) Return time distributions under variation of $\lambda_0$ from 1.04 to 1.12. Circles represent the observed distributions $P(\tau)$, and solid lines show respective power-law fits $\sim e^{\gamma \tau}$. Arrow indicates the direction of increase of $\lambda_0$. The inset shows scaling exponent $\gamma$ vs $\lambda_0$. (c) Evolution of the system's potential energy landscape $U(R)$ recovered from the data under evolution of $\lambda_0$. (d) Slices of potential energy landscape $U(R)$ for different values of $\lambda_0$. Arrow indicates the direction of increase of $\lambda_0$. (e) Fluctuations of the order parameter $R$ in the incoherent state under variation of $\lambda_0$. Circles represent medians, boxes show the interquartile ranges, and the whiskers indicate the 2.5$^{th}$ and the 97.5$^{th}$ percentiles.}
  	\label{fig4}
  \end{figure*}
%\subsubsection{Statistics of return time}
\par Thereafter, to finalize our analysis, following~\cite{frolov2022self} we explored the distribution of return times $P(\tau)$, i.e., the time intervals $\tau$ between the neighboring transitions to a coherent state (Fig.~\ref{fig4}a). We explore these distributions in the bistable domain at fixed $\alpha=0.003$ and $\beta=0.002$. Such return time distributions $P(\tau)$ were recovered from the numerically produced time courses $R(t)$ and plotted in a log-log scale in Fig.~\ref{fig4}(b) for different values of $\lambda_0$ increasing from 1.04 to 1.12 with a step of 0.01. One can see that despite the variation of control parameter $\lambda_0$, computed return time distributions $P(\tau)$ possess a linear descending part well-fitted by a power law ($p>$0.99 via the Pearsons' $\chi^2$-test for all $\lambda_0$). At smaller $\lambda_0$ signifying the entrance of the bistable domain, scaling exponent $\gamma$ is approximately -3/2 (inset in Fig.~\ref{fig4}(b)). This value of $\gamma$ is consistent with the theory of intermittent behavior at the border of synchronization~\cite{hramov2006ring} and experimental observations of return times of epileptic seizures in rodents~\cite{frolov2019statistical,koronovskii2016coexistence}. The scaling exponent of -3/2 was also reported in our previous work~\cite{frolov2022self}, where the scale-free structure of the underlying graph induced the region of SOB on the complex network. However, in the current HOI model, the scaling exponent $\gamma$ does not hold constant throughout the SOB domain but increases its negativity as the system evolves towards the forward transition (the inset in Fig.~\ref{fig4}(b)). Up to $\lambda_0=1.08$, scaling exponent $\gamma$ grows slowly and increases much faster at $\lambda_0>1.08$. It is reflected in the form of the distribution $P(\tau)$, which becomes more narrow with increasing $\lambda_0$ and even \qq{bell}-shaped (but still heavy-tailed) at $\lambda_0\geq1.1$. The form of \qq{bell} signifies an emergence of a characteristic time-scale of the transition to a coherent state, meaning that at higher $\lambda_0$ the switches between network states become more regular than spontaneous.

\par To gain more insight into the nature of this behavior we explore the potential energy landscape $U(R)$ of the system also recovered from the data (see Appendix~\ref{app:potential} to learn more about potential energy landscape construction). Fig.~\ref{fig4}(c) and (d) display the evolution of computed potential energy landscape $U(R)$ under variation of $\lambda_0$. Complementing Fig.~\ref{fig4}(e) presents the fluctuation of the order parameter $R$ as a function of $\lambda_0$. One can see that at $\lambda_0=1.04$ corresponding to $\gamma\approx-3/2$, the incoherent state of the network possesses much lower potential energy than the coherent state and is separated from the latter by a high-potential barrier. Although the system is bistable, an incoherent state is energetically much more preferable, and the system rarely escapes the potential well given the level of fluctuations provided by a relatively weak coupling. It's logical to assume that the occurrence of the system in the coherent state is indeed a rare event resulting from intermittent behavior driven by internal noise, i.e., a turbulent drift around the incoherent state $R=0$.

\par With increasing $\lambda_0$, the potentials of the barrier and the coherent state gradually decrease. The potential energy difference between the coherent state and the barrier also increases which speaks in favor of the stabilization of the coherent state. Moreover, the level of fluctuations of order parameter $R$ in the vicinity of the incoherent state also grows (Fig.~\ref{fig4}(e)) due to an increase in the available level of coupling strength in the system. These observations suggest that the coherent state becomes more accessible with increasing $\lambda_0$ explaining a shift of the return time distribution to smaller scales. Besides, facilitated internal noise (Fig.~\ref{fig4}(e)) suggests the transition to noise-induced regular rapid switches between states of the network resulting in the narrowing of the return time distributions and the development of the \qq{bell} shape.

\begin{figure}[hpt] 
  	\centerline{
  		\includegraphics[scale=0.035]{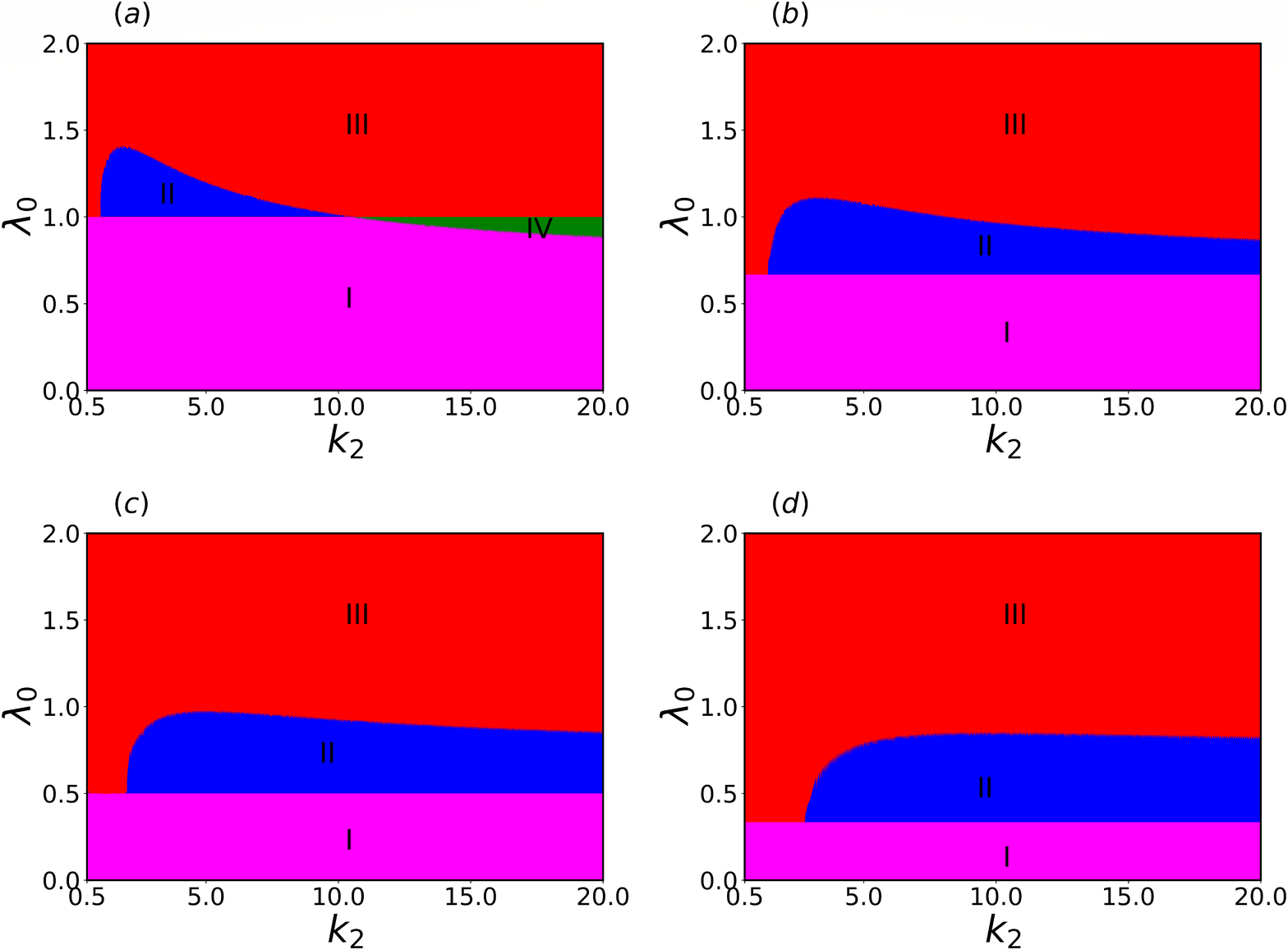}}
  	\caption{{\bf Effect of pairwise $(k_{1})$ and higher-order $(k_{2})$ coupling strengths in promoting SOB}. Phase diagram of the $(k_{2},\lambda_{0})$ parameter plane plotted upon the thermodynamic $(N \to \infty)$ limit approximation given by the Eqs. \eqref{stable_unstable_op} and \eqref{steady_lambda} for different values of pairwise coupling strength $k_{1}$: (a) $k_{1}=1$, (b) $k_{1}=1.5$, (c) $k_{1}=2$, and (d) $k_{1}=3$. Here, the regions in magenta (I), blue (II), red (III), and green (IV) are respectively the domains of subcritical, critical, supercritical, and bistable dynamics. The excitability constraint and the recovery rate are fixed at $\beta=0.002$ and $\alpha=0.002$, respectively. The figure reflects the fact that with increasing $K_{2}$, the region of critical dynamics shrinks for lower values of $k_{1}$, while for higher values of $k_{1}$, an expansion in the domain of critical dynamics becomes apparent with rising $k_{2}$.}
  	\label{k2_lambda0_space}
  \end{figure}

\par Lastly, we investigate the effect of pairwise and higher-order interactions in promoting SOB in the considered system. Specifically, we are interested in investigating the impact of the variation of pairwise and higher-order coupling strengths $k_{1}$ and $k_{2}$, respectively. The higher-order coupling strength $k_{2}$ plays a pivotal role in the initiation of a first-order transition, as recently elucidated by Skardal and Arenas \cite{skardal2019abrupt} and also predicted from our analysis. As illustrated earlier due to the presence of coupling constraint $(\beta \neq 0)$, the critical coupling $\lambda_{0}$ for forward transition shifts towards the higher value. This shift gives rise to a region of critical dynamics between the critical coupling $\lambda_{0}^{*}$ and the onset of the forward transition. Thus to investigate how the variations of $k_{1}$ and $k_{2}$ impact the critical dynamics, in Fig.~\ref{k2_lambda0_space} we plot the $(k_{2},\lambda_{0})$ parameter plane for different values of $k_{1}$: (a) $k_{1}=1$, (b) $k_{1}=1.5$, (c) $k_{1}=2$, and (d) $k_{1}=3$ with fixed excitability constraint $\beta=0.002$ and recovery rate $\alpha=0.002$. Using the thermodynamic limit approximation given by Eqs. \eqref{stable_unstable_op} and \eqref{steady_lambda} for order parameter, we characterize different regions of dynamical behaviors in the $(k_{2},\lambda_{0})$ parameter plane. The region of subcritical dynamics (I) is depicted in magenta. This region is basically the domain where $\lambda_{0}<\lambda_{0}^{*}$ and thus is independent of the variation of $k_{2}$, only the width of the area of subcritical dynamics changes with the change of $k_{1}$ since the critical transition point $\lambda_{0}^{*}$ is dependent on $k_{1}$. The region (in blue) between the onset of forward transition and the critical transition point $\lambda_{0}^{*}$ is the region of critical dynamics (II) where the coherent solution is unstable and the incoherent solution is stable. Thus, within this region, the value of the order parameter is again $r^{(1)}\approx 0$. The region (in red) beyond the onset of forward transition is the region of subcritical dynamics (III), where $0<r^{(1)}<1$. In addition to this, when the coherent solution exists even for $\lambda_{0}<\lambda_{0}^{*}$, we characterize that as the region of bistability, portrayed in green. Here, both the coherent and incoherent states are stable. For a smaller value of pairwise coupling $(k_{1}=1)$ (Fig.~\ref{k2_lambda0_space}(a)), we can observe that the interval of critical dynamics first increases with increasing $k_{2}$, however beyond a critical value of $k_{2}$, the interval of critical dynamics starts shrinking up to $k_{2} \approx 10$ and beyond this the coherent solution crosses $\lambda_{0}<\lambda_{0}^{*}$ resulting into the region of bistable dynamics. On the other hand, we can observe that as the pairwise coupling increases to higher values, the interval of critical dynamics becomes wider for larger values of $k_{2}$, and as a result, the region of bistable dynamics vanishes. It is important to notice that for $k_{2} \leq k_{1}$, there is no critical behavior in the system, as according to the coupling condition the system exhibits first-order transition only if $k_{2}>k_{1}$ and the SOB loses its meaning when the system is no longer exhibiting a first-order transition. Therefore, from these results we can conclude that the combined effect of adequate pairwise and higher-order interactions plays a pivotal role in promoting SOB in our considered model.

\section{Discussion and Conclusion} \label{conclusion}
To summarize, here we have reported a theoretical investigation of SOB exhibited by a globally coupled Kuramoto network with higher-order interactions (up to three-body interactions) while subjected to coupling constraints. Such a system mimics, for example, some basic structural properties of neuron-astrocyte coupling in the brain network, i.e., the modulation of coupling in the groups of neurons by the respective astrocyte cells. Since the dysfunction of neuron-glial interaction is believed to be a pivotal factor in epilepsy development, we suggest our simplified dynamical model can become a proper candidate to explore the role of higher-order interaction in epileptic seizure generation.

\par Our study reveals that the interplay between consumption and recovery rates of the coupling induces a delay between the critical point (where the incoherence state becomes unstable) and the forward transition to the coherent state, resulting in a region of critical bistable dynamics. Within this regime, the critical dynamics allow for a self-sustaining toggling from the state of incoherence to coherence. Notably, these spontaneous bistable switches between the incoherence and coherence state mimic the recurrence patterns seen in epileptic seizures, revealed from the power law fitting of the return time distributions with scaling exponent $-3/2$. Nonetheless, the scaling exponent doesn't maintain a consistent value across the entirety of the SOB domain. Instead, a notable trend emerges where this exponent becomes progressively more negative as the system evolves toward the forward transition to the coherent state, signifying that the switching between the coexisting states of incoherence and coherence becomes more regular than spontaneous near the forward transition. Our investigation is further accompanied by exploring the potential energy landscape of the system, which shows in proximity to the outset of the SOB domain, an incoherent state is much more preferable, and the system rarely escapes the potential well to reach a coherent state. Further, the potentials of the barrier and the coherent state steadily drop due to an increased level of internal noise induced as the strength of coupling consumption rises, suggesting that the coherent state becomes more approachable.  We also delve into the impact of pairwise and higher-order interactions in fostering SOB within our analyzed system, underscoring the essential role of synergistic pairwise and higher-order interactions in advancing SOB. Additionally, our findings reveal that no discernible SOB behavior manifests in a homogeneous, higher-ordered networked system without group interactions. This underscores the pivotal significance of higher-order interactions in our study. We note that the outcomes of our study involving a homogeneous higher-order networked system suggest that the inherent structural attributes of the network might not always dictate the onset of SOB within a networked framework. This stands in contrast to the scenario observed in the pairwise networked system, where such structural properties played a defining role \cite{frolov2022self}. This further flexibility in the choice of underlying connectivity structure can shed light on understanding the SOB theory better in networked systems approaching the first-order transition. Our theoretical model, grounded in biological inspiration, illuminates the characteristics of collective behavior that underpin the disrupting hyper-synchronization phenomenon in brain networks and opens avenues for further investigation of these phenomena within the framework of network theory.    

\bibliographystyle{apsrev4-2} % Tell bibtex which bibliography style to use
\bibliography{hoi_sob}

\appendix

\section{Results with Gaussian intrinsic frequency}
\label{app:gaussian}
\begin{figure}[hpt] 
		\centerline{
			\includegraphics[scale=0.28]{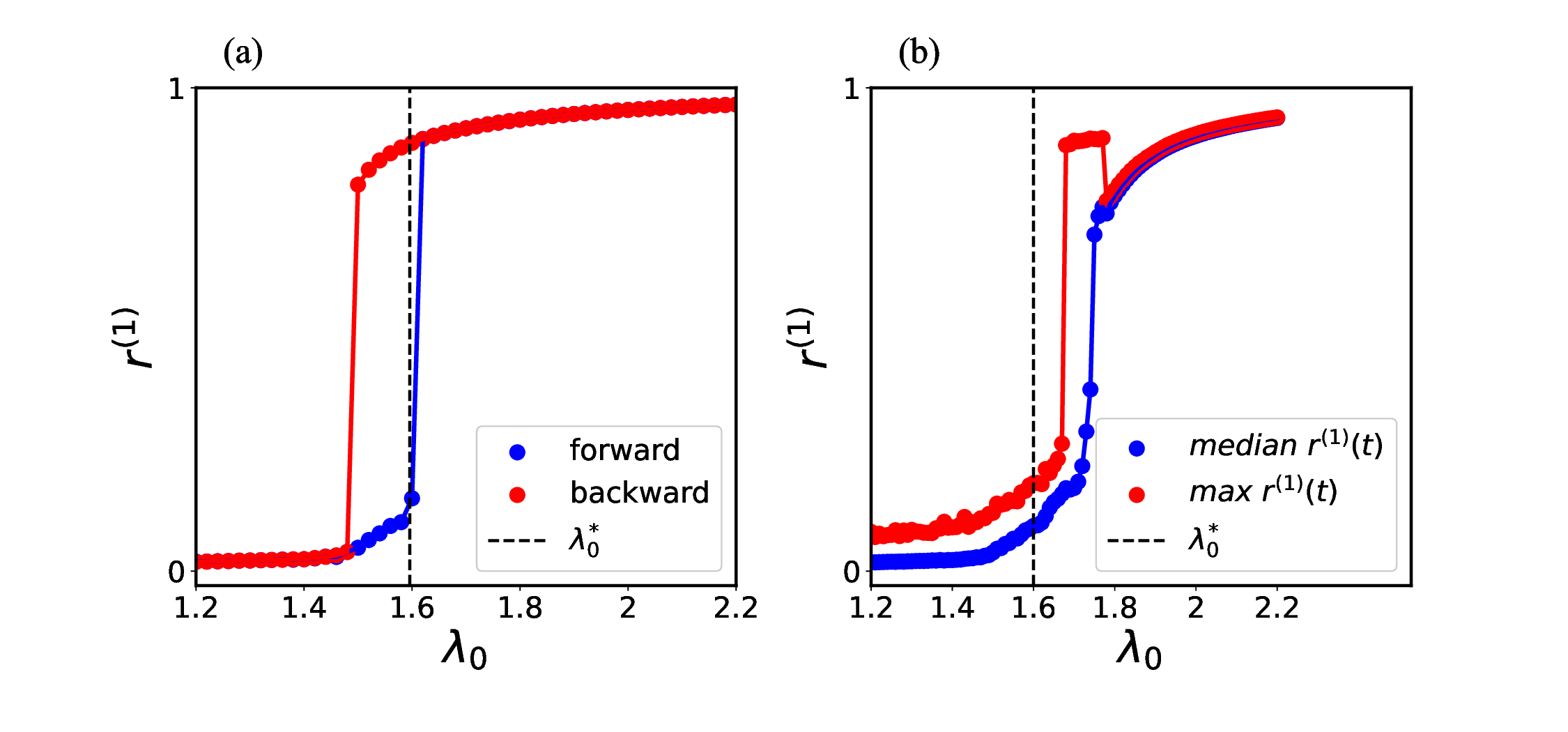}}
		\caption{{\bf Gaussian intrinsic frequency}. (a) Forward (blue circle) and backward (red circle) synchronization transition in the absence of excitability constraint $\beta $ (i.e., $\beta =0$). (b) Forward synchronization transition in the presence of excitability constraint $\beta =0.002$. The excitability recovery rate for both panels is fixed at $\alpha =0.005$. In (a), the forward and backward synchronization diagram and in (b), the maximum (red circle) and median value (blue circle) of order parameter $r^{(1)}$ are evaluated over a long time interval of $3\times 10^{4}$ time units after an initial transient period of $10^{3}$ time units. The dashed vertical black lines in (a) and (b) represent the critical synchronization transition $\lambda^{*}_{0}$, given by Eq. \eqref{critical_point}. }
		\label{normal_beta0_beta002}
	\end{figure}

Here, we consider the intrinsic frequencies of each oscillator to be drawn from the standard Gaussian distribution. The pairwise and triadic coupling strengths are taken to be $k_{1}=1$ and $k_{2}=1$, respectively, for which the system exhibits an abrupt transition to the coherent state [see Fig. \ref{normal_beta0_beta002} (a)]. While under coupling consumption (i.e., in the presence of excitability constraint $\beta =0.002$, similar to the case of Lorentzian frequency distribution, we can observe that the forward transition to the coherent state is significantly delayed with respect to the critical transition point $\lambda_{0}^{*}$ [see Fig. \ref{normal_beta0_beta002}(b)]. 	

  \begin{figure}[hpt] 
     	\centerline{
     		\includegraphics[scale=0.35]{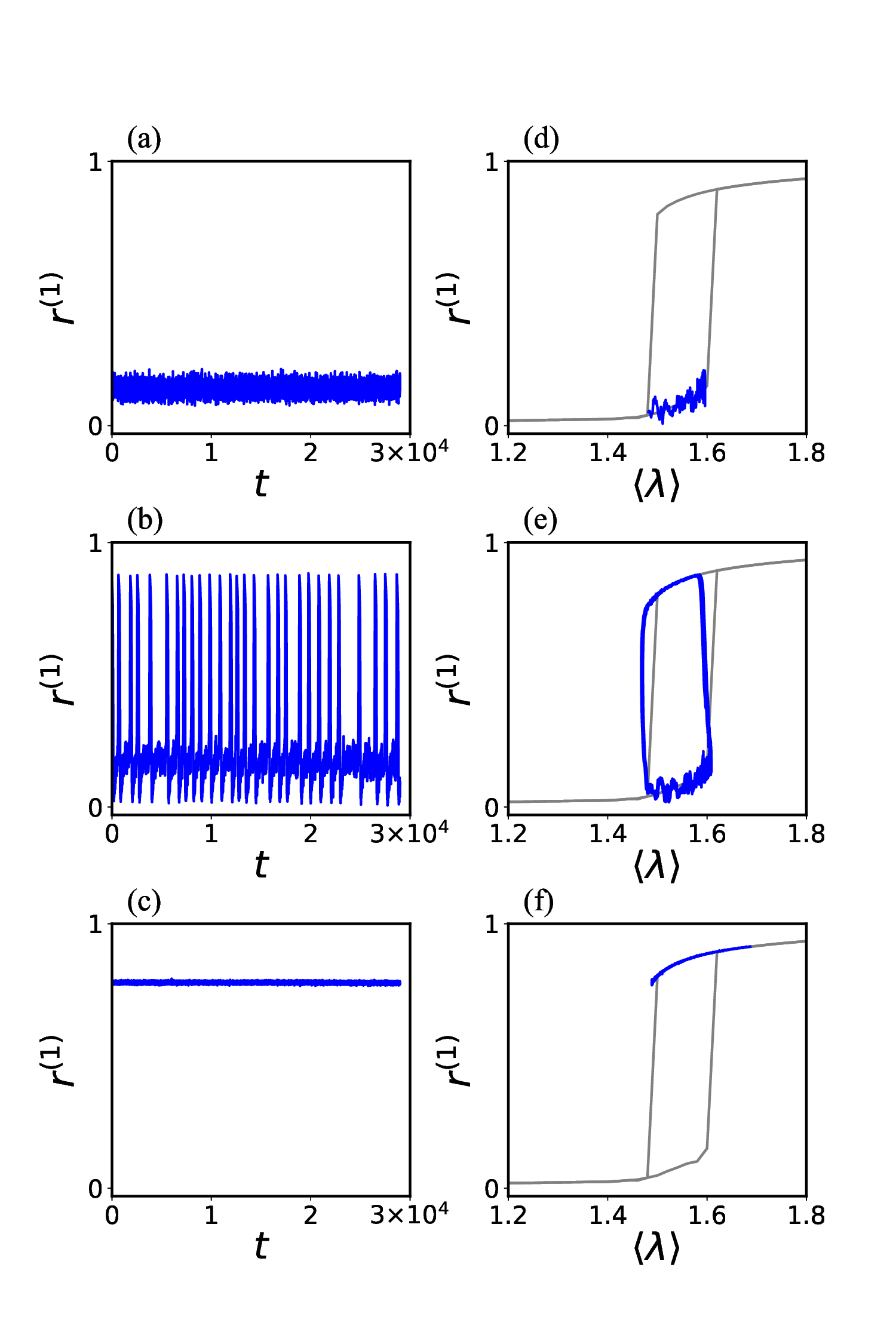}}
     	\caption{{\bf Gaussian intrinsic frequency}. Macroscopic dynamics of higher-order globally coupled network. The left and right columns represent the long-term time dependency of global order parameter $r^{(1)}$ and corresponding phase portrait on the $(r^{(1)},<\lambda>)$ plane respectively, for fixed values of $\alpha =0.005$, $\beta =0.002$ and different excitability bath depth $\lambda_{0}$. In the top, middle, and bottom rows, the value of excitability bath depth is respectively, $\lambda_{0}=1.65$, $1.68$, and $1.8$.}
     	\label{normal_ts}
     \end{figure}
\par Depending on the value of excitability recovery rate $\alpha $, and excitability constraint $\beta $, in this scenario also we can identify three distinct dynamic regions (i.e., region of subcritical dynamics, critical bistable domain and the domain of supercritical dynamics). In Fig. \ref{normal_ts}, we plot the time-series of order parameter $(r^{(1)}(t)$, accompanied by the phase portrait on the $(r^{(1)}, \langle \lambda \rangle)$ plane for typical values of $\lambda_{0}$, which illustrates these distinct dynamics of the higher-order model.
    
\par
\begin{figure}[hpt] 
       	\centerline{
       		\includegraphics[scale=0.26]{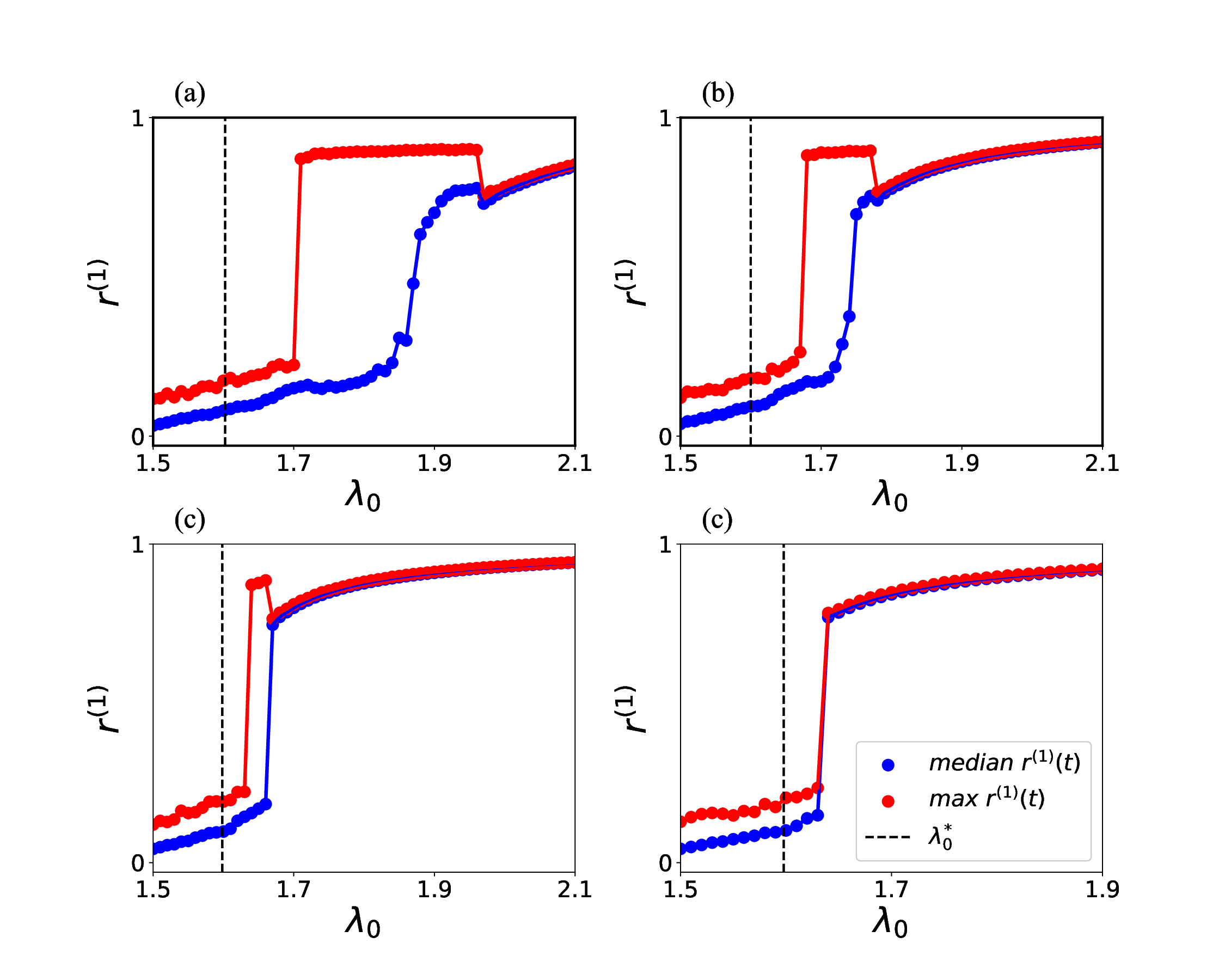}}
       	\caption{{\bf Gaussian intrinsic frequency}. Forward synchronization transition in terms of $r^{(1)}(\lambda_{0})$ for fixed value of excitability constraint $\beta =0.002$ and different values of excitability recovery rate $\alpha $: (a) $\alpha =0.003$, (b) $\alpha =0.005$, (c) $\alpha =0.008$, and (d) $\alpha =0.01$, respectively. }
       	\label{normal_alpha_variation}
       \end{figure}

 Furthermore, we also show that the system is less affected by the impact of coupling constraints when the value of recovery rate $\alpha $ is being increased keeping the consumption rate fixed at a typical value $\beta =0.002$. Figure \ref{normal_alpha_variation} shows the corresponding result, reflecting that with increasing $\alpha $, the region of critical bistable dynamics is diminished and the system resembles its original dynamics in the absence of coupling constraint. However, unlike the Lorentzian frequency distribution, here we cannot find the expressions for the coherent state in the thermodynamic limit.          

	%%%	\section*{Data availability} The data corresponding to the real-world
	%example is publicly available at \url{http://www.sociopatterns.org}. 
	
\section{Potential energy landscape reconstruction}
\label{app:potential}

With this aim, we suggest that the time course $R(t)$ is a one-dimensional process described by the stochastic differential equation:
\begin{equation}
\label{potential_stoch}
    dR(t) = -U'(R, t)dt + DdW(t),
\end{equation}
where $U(t)$ is a potential energy, $D=\sigma^2(R)/2$ is a diffusion constant and $W(t)$ is a Wiener process driving evolution of $R(t)$. One way to solve Eq.~(\ref{potential_stoch}) equation is to introduce a probability $p(R,t)$ and to write a corresponding Fokker-Planck equation:
\begin{multline}
\label{potential_FP}
    \frac{\partial}{\partial t} p(R,t) = \frac{\partial}{\partial R} \left[ U'(R,t) p(R,t) \right] \\ + \frac{\partial^2}{\partial R^2}\left[ D(t) p(R,t) \right].
\end{multline}
Seeking for a stationary solution of Eq.~(\ref{potential_FP}), we assume $\partial p(R,t)/\partial t=0$. Therefore:
\begin{equation}
\label{stationary_FP}
    U(R) p(R) = -D\frac{\partial}{\partial R}\left[ p(R,t) \right],
\end{equation}
solving which one finds:
\begin{equation}
\label{stationary_solution}
    U(R)/D = - \log\left[p(R)\right].
\end{equation}

According to Hirota et al.~\cite{hirota2011global} and Curtin et al.~\cite{curtin2020dysregulated}, the potential energy is presented in the normalized units $U(R)/\sigma^2(R)$.

\end{document}